\documentstyle[psfig,onecolumn]{mn2e}
\newif\ifAMStwofonts

\newcommand{\etal}{{et al.} }
\newcommand{\adhoc}{{ad hoc} }

\newcommand{\astroh}{{\it Astro-H} }
\newcommand{\astrohp}{{\it Astro-H}}
\newcommand{\mytorus}{{\sc MYTorus} }
\newcommand{\mytorusp}{{\sc MYTorus}}

\newcommand{\mchyphen}{{Monte Carlo} }

\newcommand{\chandra}{{\it Chandra} }

\newcommand{\feka}{{Fe~K$\alpha$} }

\newcommand{\fekalfa}{{Fe~K$\alpha$} }

\newcommand{\fekb}{{Fe~K$\beta$} }

\newcommand{\fekbeta}{{Fe~K$\beta$} }

\newcommand{\nh}{$N_{\rm H}$ }
\newcommand{\nhp}{$N_{\rm H}$}

\newcommand{\thetaobs}{{$\theta_{\rm obs}$} }
\newcommand{\thetaobsp}{{$\theta_{\rm obs}$}}

\newcommand{\tablecosrange}{{Table~1} }

\newcommand{\figcsanatomy}{{Fig.~1} }
\newcommand{\figcsanatomyp}{{Fig.~1}}
\newcommand{\figfllcsratio}{{Fig.~2} }
\newcommand{\figfllcsratiop}{{Fig.~2}}
\newcommand{\figfekmcshoulders}{{Fig.~3} }
\newcommand{\figfekmcshouldersp}{{Fig.~3}}
\newcommand{\figfekeshoulders}{{Fig.~4} }
\newcommand{\figfekeshouldersp}{{Fig.~4}}
\newcommand{\figfekcsdifgamm}{{Fig.~5} }

\newcommand{\figfekvbroadfour}{{Fig.~6} }

\title{The Compton shoulder of the Fe~K$\alpha$ 
fluorescent emission line in active galactic nuclei}

\author[Tahir Yaqoob \& Kendrah D. Murphy]
{Tahir Yaqoob$^{1}$ and Kendrah D. Murphy$^{2,3}$ \\
$^{1}$Department of Physics and Astronomy, Johns Hopkins University, Baltimore, MD 21218. \\
$^{2}$MIT Kavli Institute for Astrophysics and Space Research, 77 Massachusetts
        Avenue, NE80-6013, Cambridge, MA 02139. \\
$^{3}$Department of Physics, Skidmore College, 815 North Broadway, Saratoga Springs, NY 12866. \\
}

\date{Accepted. Received; in original form}

\begin{document}

\maketitle

\begin{abstract} 

We present new, high signal-to-noise ratio
results from a Monte Carlo study of the properties of the
Compton shoulder of the \fekalfa emission line in the toroidal X-ray 
reprocessor model of Murphy \& Yaqoob (2009). The model is valid
for equatorial column densities in the
range $10^{22} \ \rm cm^{-2}$ to $10^{25} \ \rm cm^{-2}$,
which comprehensively covers the
Compton-thin to Compton-thick regimes. 
We show how
the shape of the Compton shoulder and its flux relative to the 
core of the \fekalfa emission line depend on the
torus column density and orientation, for the case
of a half-opening angle of $60^{\circ}$ and cosmic abundances.
The variety of Compton shoulder
profiles is greater than that for both (centrally-illuminated) spherical 
and disk geometries. Our Monte Carlo simulations
were done with a statistical accuracy that is
high enough to reveal, for the case of an edge-on, Compton-thick 
torus, a new type of Compton shoulder that is not present in the
spherical or disk geometries. Such a Compton shoulder is dominated by a
narrow back-scattering feature peaking at $\sim 6.24$~keV.
Our results are also sensitive enough 
to reveal a dependence of the
shape of the Compton shoulder 
(and its magnitude relative to the \fekalfa line core) 
on the spectral shape of the incident X-ray continuum.
We also present results of the 
effect of velocity broadening on the \fekalfa line
profile and find that if either the velocity width or instrument resolution
is greater than a FWHM of $\sim 2000 \ \rm km \ s^{-1}$, the Compton shoulder
begins to become blended with the line core and the characteristic features
of the Compton shoulder become harder to resolve. In particular, at 
a FWHM of $\sim 7000 \ \rm km \ s^{-1}$ the Compton shoulder is not resolved
at all, its only signature being a weak asymmetry in the blended line profile.
This means that CCD X-ray detectors cannot 
unambiguously resolve the Compton shoulder.
Our results are freely available in a format that is suitable for
direct spectral-fitting of the continuum and line model to real data.

\end{abstract}


\section{Introduction}
\label{csintro}

The narrow \fekalfa emission line in Active Galactic Nuclei (AGNs)
is by now an established feature of the X-ray spectrum of both type~1 and type~2
sources (e.g. see Shu, Yaqoob, \& Wang 2010, and references therein).
The FWHM of the emission line is usually found to be less than $\sim 5000 \ \rm \ km \ s^{-1}$ or so,
indicating an origin in the outer (optical) broad line region (BLR) or further out,
in the putative parsec-scale torus (e.g. Yaqoob \& Padmanabhan 2004; Bianchi \etal 2008;
Shu \etal 2010). From a subsample of the
highest quality \chandra HETG spectra, Shu \etal (2010) in fact found
that the width of the \fekalfa line relative to the optical line width of
H$\beta$ varies from source to source so that there may not be a ``universal'' location
for the X-ray reprocessor.
X-ray spectroscopy also shows overwhelming evidence for the \fekalfa line peaking at
$\sim 6.4$~keV, indicating that the matter responsible for producing this
line is essentially neutral (e.g. Sulentic
\etal 1998; Weaver, Gelbord, \& Yaqoob 2001; Reeves 2003; Page \etal 2004; Yaqoob \& Padmanabhan 2004;
Jim\'{e}nez-Bail\'{o}n \etal 2005;  Zhou \& Wang 2005; Jiang, Wang, \& Wang 2006; Levenson \etal 2006;
Shu \etal 2010). Although 
emission lines from ionized species of Fe are observed in some AGN
(e.g. Yaqoob \etal 2003; Bianchi \etal 2005, 2008), the present paper is
concerned specifically with the \fekalfa line component that is
centered around 6.4~keV.  
Although some type~2 AGN do not show evidence
for line-of-sight X-ray obscuration (Brightman \& Nandra 2008), in general
the ``type classification'' of AGN
appears to be related to the orientation of the structure
of circumnuclear matter in the central engine. This, along with the fact that
the  narrow \fekalfa line at $\sim 6.4$~keV appears in the  X-ray spectra of AGN regardless
of whether or not the X-ray spectrum shows line-of-sight obscuration
suggests that the material producing the \fekalfa emission line is
in some sense toroidal.

The modification of the intrinsic
X-ray spectrum due to the effects of absorption, scattering and fluorescence in
the circumnuclear matter are not trivial to calculate in the Compton-thick
regime since the reprocessing is a complicated function of
column density,
geometry (and covering factor), element abundances, and the system inclination angle
(e.g. Ghisellini, Haardt, \& Matt 1994; Ikeda, Awaki, \& Terashima 2009).
To obtain the most accurate spectra, Monte Carlo simulations are required.
In Murphy \& Yaqoob (2009, hereafter MY09) and Yaqoob \etal (2010) we presented the results
of Monte Carlo calculations for the reprocessed X-ray continuum and the \fekalfa
emission line in a toroidal geometry, valid for column densities
in the range $10^{22}$ to $10^{25} \ \rm cm^{-2}$, thus 
covering the Compton-thin to Compton-thick regimes. Only the results for
the zeroth-order, or core, component for the \fekalfa emission line in the MY09
model have been presented thus far (the zeroth-order component corresponds
to line photons that escape the medium without scattering). The scattered
emission-line photons constitute the so-called ``Compton shoulder'', which
can provide additional critical information on the column density
and orientation of the line emitter (e.g.
Ghisellini \etal 1994; Sunyaev \& Churazov 1996;
Matt 2002; Watanabe \etal 2003).
Given the high degree of degeneracy in modeling complex X-ray spectra,
including a physical model of the Compton shoulder (as opposed to
a simple \adhoc component) is important, especially in view of the
fact that even the non-detection of a Compton shoulder provides constraints
on the physical parameters of the model. Quantifying upper limits
on the Compton shoulder when it is not detected particularly
demands a physically-motivated model. We have substantially improved
the statistical accuracy of the X-ray spectra of the toroidal 
reprocessor model (\mytorusp) described in MY09 in order
for it to be suitable for spectral-fitting to real data.
The Monte Carlo results have been put into formats that are
readily useable in conjunction with standard X-ray spectral-fitting
packages and we have made the \mytorus model publicly
available\footnote{See http://www.mytorus.com}.
In the present paper we describe the detailed form of the Compton shoulder
of the \fekalfa line in a toroidal geometry and discuss its dependence
on the model parameters.

We note that the \mytorus model is not restricted to any absolute size scale so it
can be applied to {\it any} toroidal distribution of matter that
is centrally-illuminated by X-rays.
Gaskell, Goosmann, \& Klimek (2008) argue that there is considerable
observational evidence that the BLR itself has a toroidal
structure, and that there may be no distinct boundary between the
BLR and the classical parsec-scale torus.
Gaskell \etal (2008) also argue that there may even be no distinction
between the outer accretion disk and the BLR. A toroidal
distribution of matter may exist anywhere from the outer accretion
disk to parsec-scale distances from the X-ray continuum source.
Thus, throughout the present paper, we shall refer to {\it ANY} toroidal
distribution of matter in the central engine as ``the torus'',
regardless of it's actual size or physical location in the AGN central engine.
Our torus model complements hydrodynamical, photoionized {\it torus wind} simulations 
such as those of Dorodnitsyn \& Kallman (2009) which are designed to
account for the emission lines from ionized species in AGN, and the
photoionized absorber and emission lines from ionized in some type~1 AGN.
Such models of evaporation of the inner torus do not account for the
\fekalfa line component that is centered around $\sim 6.4$~keV, which
must originate in the cooler, unionized region of the torus.

The paper is organized as follows.
In \S\ref{mytorusmodel}
we give a brief overview of the assumptions of the \mytorus model
and in \S\ref{linecsdef} we
clarify the definition of the \fekalfa line Compton shoulder in the present paper.
In \S\ref{fllcsratio} we
present the results for the ratio of the 
\fekalfa line Compton shoulder flux to the zeroth-order flux, and
in \S\ref{fllcspardepend} we discuss the
characteristic shapes of the Compton shoulder profile as a function
of some of the model parameters.
In \S\ref{csgamdepend} we illustrate the dependence of the shape of
the Compton shoulder profile on the form of the intrinsic,
incident X-ray continuum, and in 
\S\ref{fllvelbroad} we discuss the effects of velocity broadening
on the \fekalfa line core and Compton shoulder.
We summarize our conclusions in \S\ref{summary}.

\section{Overview of Model Assumptions}
\label{mytorusmodel}

Here we give a brief overview of the critical
assumptions that the model Monte Carlo simulations are  based upon.
Full details can be found in MY09.
Our geometry is an azimuthally-symmetric
doughnut-like torus with a circular cross-section,
characterized by only two parameters, namely
the half-opening angle, $\theta_{0}$, and the
equatorial column density, \nh (see Fig.~1 in MY09).
We assume that the X-ray source
is located at the center of the torus and emits isotropically and that
the reprocessing material is uniform and
essentially neutral (cold).
For illumination by an X-ray source that is emitting isotropically,
the mean column density, integrated over all incident
angles of rays through
the torus, is $\bar{N}_{H}$~$=(\pi/4)$\nhp.
The inclination angle between the observer's line of sight and the
symmetry axis of the torus is denoted by
\thetaobsp, where \thetaobs$=0^{\circ}$ corresponds to a face-on
observing angle and \thetaobs$=90^{\circ}$
corresponds to an edge-on observing angle. 
In our calculations we distribute the emergent photons in
10 angle bins between $0^{\circ}$ and $90^{\circ}$ that have
equal widths in $\cos{\theta_{\rm obs}}$, and refer
to the face-on bin as \#1, and the edge-on bin as \#10
(see \tablecosrange in MY09).

The value of $\theta_{0}$ for which we have calculated
a comprehensive set of models is $60^{\circ}$,
for $N_{H}$ in the range $10^{22} \ \rm cm^{-2}$ to 
$10^{25} \ \rm cm^{-2}$, valid 
for input spectra with energies in the range 
0.5--500~keV (see MY09 for details). For
$\theta_{0}=60^{\circ}$, the solid angle subtended by the torus at the
X-ray source, $\Delta\Omega$, is $2\pi$
(and we refer to $[\Delta\Omega/(4\pi)]$ as a covering factor,
which in this case is 0.5). The covering factor may also be
expressed in terms of the physical 
dimensions of the torus. If $a$ is the
radius of the circular cross-section of the torus, and $c+a$ is
the equatorial (i.e. maximum) 
radius of the torus then $[\Delta\Omega/(4\pi)]=(a/c)$
(see MY09).
Our model employs a
full relativistic treatment of Compton scattering,
using the full differential and total Klein-Nishina Compton-scattering
cross-sections.
We utilized photoelectric absorption cross-sections for 30 elements as
described in Verner \& Yakovlev (1995) and
Verner \etal (1996) and
we used Anders and Grevesse (1989) elemental cosmic abundances in
our calculations.
The Thomson depth 
may also be expressed in terms of the column density:
$\tau_{\rm T} = K N_{\rm
H}\sigma_{\rm T} \sim 0.809N_{24}$ where $N_{24}$ is the column density in
units of $10^{24} \rm \
cm^{-2}$.
Here, we have employed
the mean number of electrons per H atom, $\frac{1}{2}(1+\mu)$,
where $\mu$ is the mean molecular weight.
With the abundances of Anders \& Grevesse, $K=1.21656$, assuming that the number of
electrons from all other elements aside from H and He is
negligible.
For the \fekalfa and \fekbeta fluorescent emission lines,
we used rest-frame line energies of 6.4008~keV and 7.058~keV
respectively, appropriate for neutral matter (e.g.
see Palmeri \etal 2003).
We used a fluorescence yield for Fe of 0.347
(see Bambynek \etal 1972) and an \fekb to \feka
branching ratio of 0.135
(representative of the range of
experimental and theoretical values discussed in Palmeri \etal 2003,
see also Kallman \etal 2004).

Compared to MY09, the results in the present paper have
a substantially higher statistical accuracy 
because they are based
on Monte Carlo simulations with higher numbers of
injected rays at each energy, and the calculations
employ the method of
weights (as opposed to following individual photons).
The statistical accuracy actually achieved depends on the 
model parameters. The most challenging regime corresponds to
the highest column density torus ($10^{25} \ \rm cm^{-2}$) observed
edge-on, and in this worst-case scenario the 
statistical errors on the Compton shoulder profile (for the
default torus opening angle and abundances) are better than
$2\%$ per 100~eV energy bin. On the other hand,
for a face-on torus with a
column density of $10^{24} \ \rm cm^{-2}$, 
the corresponding statistical errors on the Compton
shoulder profile are better than $0.2\%$ per 100~eV bin.

Throughout the present paper we present results for power-law
incident continua (in the range 0.5--500~keV), characterized
by a photon index, $\Gamma$, by integrating
the basic monoenergetic Monte Carlo results (Greens functions--
see MY09).

\section{Definition of The \feka Line Compton Shoulder}
\label{linecsdef}

The zeroth-order component of the \fekalfa fluorescent emission line
refers to line photons that escape without any
interaction with the medium that they were
created in. Although they are emitted by atoms/ions isotropically,
the angular distribution of the emerging zeroth-order line
photons in general depends on the geometry, unless the
medium is optically-thin to the line photons. The zeroth-order line
photons constitute the majority of photons in the emergent emission
line for the parameters pertaining to our model assumptions, 
and they all have the same energy
in the \mchyphen results
because they have not interacted with the medium. When velocity
broadening is applied to the zeroth-order line emission the
photon energies are of course modified to reflect the
distribution of the broadening function. Most of the
fluorescent \fekalfa
emission-line measurements in the literature that have been modeled
with simple Gaussian functions correspond to the zeroth-order of
the emission line. However, depending on the spectral resolution
of the instrument, some of the measured flux of this
line core may include a
contribution from the Compton-scattered line photons (or Compton
shoulder) since the scattered line component consists of photons
with a range of energies going all the way up to the
zeroth-order line energy. 
If one compares the results of applying the \mytorus model
to results in the literature that were obtained using \adhoc models
for the \fekalfa line core and Compton shoulder
one must scrutinize precisely which components
of the line were actually measured with the \adhoc model components.

We refer to fluorescent emission-line photons that escape the
medium after at least one interaction
as the scattered component of the line.
Compton scattering shapes the energy distribution of the scattered
line photons. However, the final scattered line profile depends
on the geometry, orientation, and column density distribution of the
medium because the escape probabilities after scattering may be
highly directional. In cold matter,
the scattered photon distribution
resulting from the $n{\rm th}$ scattering has a spread in energy
from $E_{0}$ (the rest-frame zeroth-order line energy), down to
$511/[(511/E_{0} \ {\rm keV})+ 2n]$~keV. The number of photons
in each scattering order is diminished compared to the number in the
previous scattering order. The exact dependence of the relative number
of photons in each scattering, and therefore of the measurable
width of the scattered distribution, is a function of 
model parameters. If the optical depth to absorption of
the line photons is much greater or much less than unity, only the
first scattering may dominate. Even for intermediate optical
depths the third scattering is, for practical
purposes, negligible in flux compared to the
first scattering.

\begin{figure}
\centerline{
\psfig{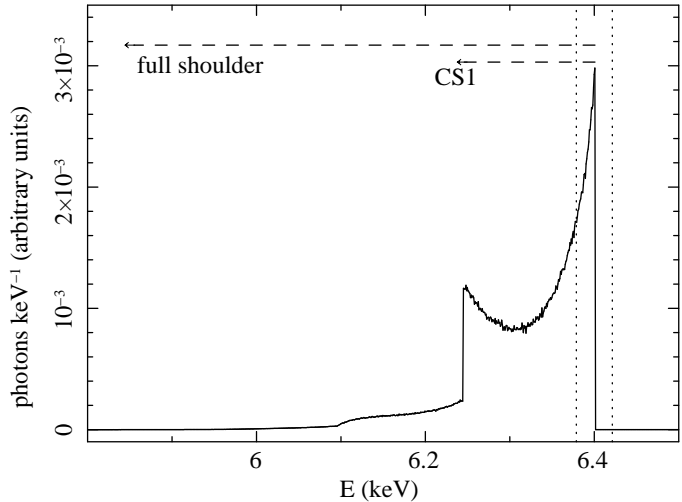}}
\caption[Definition of Compton Shoulder.]{\footnotesize
Illustration of our definition of the Compton shoulder for
an emission line, compared with the definition ``CS1''. Our
definition (``full shoulder'' in the figure)
includes ALL of the scattered line flux, but ``CS1''
includes all of the flux only in the energy interval of the first scattering.
Note that no velocity broadening has been applied.
When velocity
broadening is applied, \adhoc models will confuse line flux
from the zeroth-order and the Compton shoulder (regardless of
definition). The vertical dotted lines in the figure illustrate
this by showing the energy width corresponding to
$\pm 1000 \ \rm km \ s^{-1}$ either side of the zeroth-order
line energy. 
}
\end{figure}

The term ``Compton shoulder'' has been used in the
literature in more than one way. The most common usage refers to the
line flux between the energy extrema of the first scattering,
{\it which includes some contributions from all scatterings}. This is
because the contributions from all
scattering orders extends to the zeroth-order line energy and
different scattering orders
cannot be measured separately. This is referred to as ``CS1''
(e.g. see Matt 2002),
and is illustrated in \figcsanatomyp.
Some literature may refer to only the first
scattering as the Compton shoulder, but this of course can only have
a theoretical context. A third definition, which is the one we
use here and throughout the present paper, is that the Compton shoulder
includes {\it all} of the scattered line flux, for all scatterings
(see \figcsanatomyp).
This is the most appropriate definition for a model such as \mytorus
which self-consistently calculates the zeroth-order and the
entire scattered
line profile and does not allow the relative fluxes of the two
components to vary in an \adhoc fashion. Definitions such as CS1
are not necessary for our purpose because there is no need to fit the scattered
component of an emission line separately.

In practice it may not actually be possible to observationally
distinguish
the zeroth-order component of an emission line from its Compton
shoulder. The finite energy resolution of the instrument and/or the
velocity broadening (of all the emission-line components)
may confuse the two blended components of a line. This is illustrated in
\figcsanatomy with dotted lines placed at
energies corresponding to $\pm 1000 \ \rm km \ s^{-1}$
either side of the zeroth-order line energy. Having said that,
one will never need to observationally distinguish between the
zeroth-order component of an emission line from its Compton shoulder
with the \mytorus model because it is a self-consistent model.
However, it is important to be aware of the ``mixing'' for
the interpretation of spectral-fitting results, and more will be said
on this topic when we discuss velocity broadening in \S\ref{fllvelbroad}.

\section{Relative Magnitude of the Compton Shoulder}
\label{fllcsratio}

The relative magnitude of the zeroth-order and scattered
line components is fixed by the physics and geometry and 
in \figfllcsratio we show the ratio of the \fekalfa line Compton shoulder
flux to the zeroth-order line flux (hereafter, the CS ratio)
as a function of $N_{H}$ for two different
values of \thetaobs (corresponding to the face-on and edge-on
inclination angle bins --see \tablecosrange in MY09).
The calculations in \figfllcsratio were made for an
incident power-law continuum with a photon index of 1.9.
It can seen that the ratio peaks at $N_{H} \sim 2-3
\ \times 10^{24} \ \rm cm^{-2}$,
reaching a maximum of $\sim 0.29$ (face-on),
and $\sim 0.37$ (edge-on).
For the face-on case, the CS ratio remains at $\sim 0.29$
for higher column densities since
the Compton shoulder photons escape from 
within a Compton-depth or so from the illuminated surfaces
of the torus for lines-of-sight that are not obscured.
For the edge-on case the CS ratio declines as function
of column density after reaching its maximum value, due to a higher
probability of absorption at higher column densities.
In the optically-thin limit, one might expect the
CS ratio to be approximately $\sim 
K\sigma_{T} (\pi/4)N_{H} = 0.809(\pi/4)N_{24}$, and this 
relation is shown in \figfllcsratio (dotted
line). The actual CS ratios fall below this line because the 
\fekalfa line photons are formed throughout the medium, so that
the average column density for scattering of the line photons
is less than $(\pi/4)N_{H}$. At $N_{H}=10^{22} \ \rm cm^{-2}$,
the CS ratio is a factor $\sim 0.80$, and $\sim 0.87$ less than
$0.809(\pi/4)N_{24}$ for the face-on and edge-on orientation respectively.
The face-on orientation obviously has a smaller mean column density
than the edge-on one.

Matt (2002) showed the CS ratio for spherical and disk geometries
(albeit for the ``CS1'' definition of the Compton shoulder).
The spherical geometry shows a maximum CS ratio of $\sim 0.42$
for $N_{H} \sim 2 \ \times 10^{24} \ \rm cm^{-2}$, 
similar to the case for the edge-on torus (and note that
the spherical radial column densities should be compared
with $\sim (\pi/4)$ times our
equatorial torus column densities).
The Compton-thick disk geometry shows a lower CS ratio, peaking at $\sim 0.2$
(Matt 2002), less than the corresponding value for the Compton-thick
face-on torus. The range of incident and emergent
photon angles are very different for the
two geometries and the differences in the reflection
spectra from the torus and disk geometries have been discussed in MY09.

We found that for all values of \thetaobs for the torus,
there was no detectable difference in the CS ratio as a function
of $\Gamma$ up to the \nh value that gives the maximum CS ratio
(for a given value of \thetaobsp). After that, the CS ratios
diverge for different values of $\Gamma$, with flatter incident
continua giving larger CS ratios. For the face-on case,
the CS ratio at $N_{H} = 10^{25} \ \rm cm^{-2}$ varies
between $\sim 0.30$ to $\sim 0.32$ as $\Gamma$ varies from
2.5 to 1.5. For the edge-on case, the CS ratio at $N_{H} = 10^{25} \ \rm cm^{-2}$ varies
between $\sim 0.28$ to $\sim 0.27$ as $\Gamma$ varies from
2.5 to 1.5. Flatter spectra have relatively more continuum photons
at higher energies so that the \fekalfa line photons
are produced deeper in the medium, increasing the average
Thomson depth for zeroth-order line photons to scatter before escaping.
Although these variations in the CS ratio with $\Gamma$ are small,
the effect of different values of $\Gamma$ on the Compton shoulder
{\it shape} is much more pronounced (see \S\ref{csgamdepend}).

\begin{figure}
\centerline{
\psfig{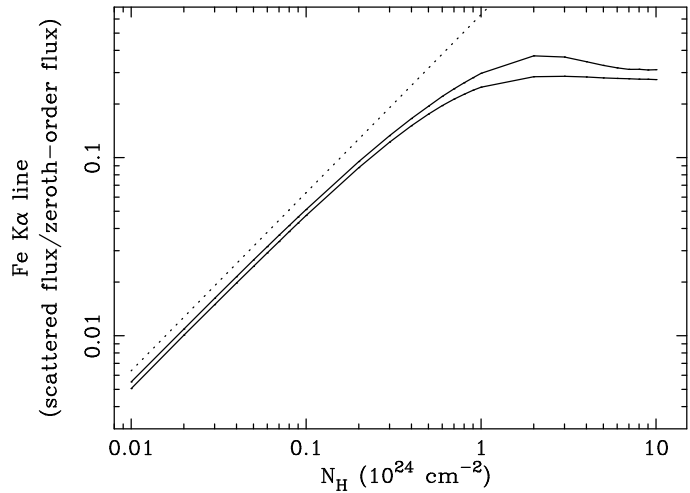}}
\caption[Ratio of the \fekalfa line Compton shoulder flux to the
zeroth-order core flux.]{\footnotesize
The ratio of the scattered flux to the zeroth-order flux in
the \fekalfa line from the \mytorus model, as a function of
\nh (for $\Gamma=1.9$). Shown are curves for the
face-on (lower curve) and edge-on (upper curve) inclination-angle bins.
The dotted curve corresponds to a CS ratio of
$0.809(\pi/4)N_{24}$ in order to compare with the Monte Carlo results in
the optically-thin limit.
See text in \S\ref{fllcsratio} for further details.
}
\vspace{-3mm}
\end{figure}

\section{The Shape of the Compton Shoulder}
\label{fllcspardepend}

The shape of the Compton shoulder of a fluorescent
emission-line escaping from the torus has a dependence on
the column density and inclination angle of the torus.
It is also affected by the shape of the incident continuum
spectrum, the covering factor (or opening angle), and
element abundances. In this section we discuss the shape of the
\fekalfa line shoulder for the default covering factor ($[a/c]=0.5$)
and solar abundances.
\figfekmcshoulders and \figfekeshoulders
illustrate the shapes of the Compton shoulder for a power-law
incident continuum with $\Gamma=1.9$, for four column densities
($10^{23}$, $10^{24}$, $5 \times 10^{24}$, and
$10^{25} \rm \ cm^{-2}$) and different inclination angles of the torus.
\figfekmcshoulders and \figfekeshoulders show Compton
shoulders for the same physical parameters
except that in the former the profiles are plotted against
Compton wavelength shift and in the latter they are
plotted against absolute energy.
The Compton shoulders in \figfekmcshoulders 
are in units of the {\it dimensionless} Compton
wavelength shift with respect to the zeroth-order rest-frame
energy of the emission line. In other words, if $E$ is the energy
of a line photon, and $E_{0}$ is the zeroth-order line energy,
$\Delta \lambda = (511 \ {\rm keV}/E)-(511 \ {\rm keV}/E_{0})$.
The Compton shoulder shapes shown in 
both \figfekmcshoulders and \figfekeshoulders are
from the original
\mchyphen results and have no velocity broadening applied to them. 
In order to facilitate a direct comparison of the Compton
shoulder profile shapes for different column densities and 
inclination angles, all of the profiles in \figfekmcshoulders  have
been normalized to a flux of unity. It should be remembered that
the absolute flux of the Compton shoulder varies
significantly with column density, and the flux ratio for
two column densities can be estimated using \figfllcsratiop.
The Compton shoulder profiles that are plotted against
energy (\figfekeshouldersp) {\it do} show {\it relative}
fluxes for different parameters that reflect the original
Monte Carlo results.

\begin{figure}
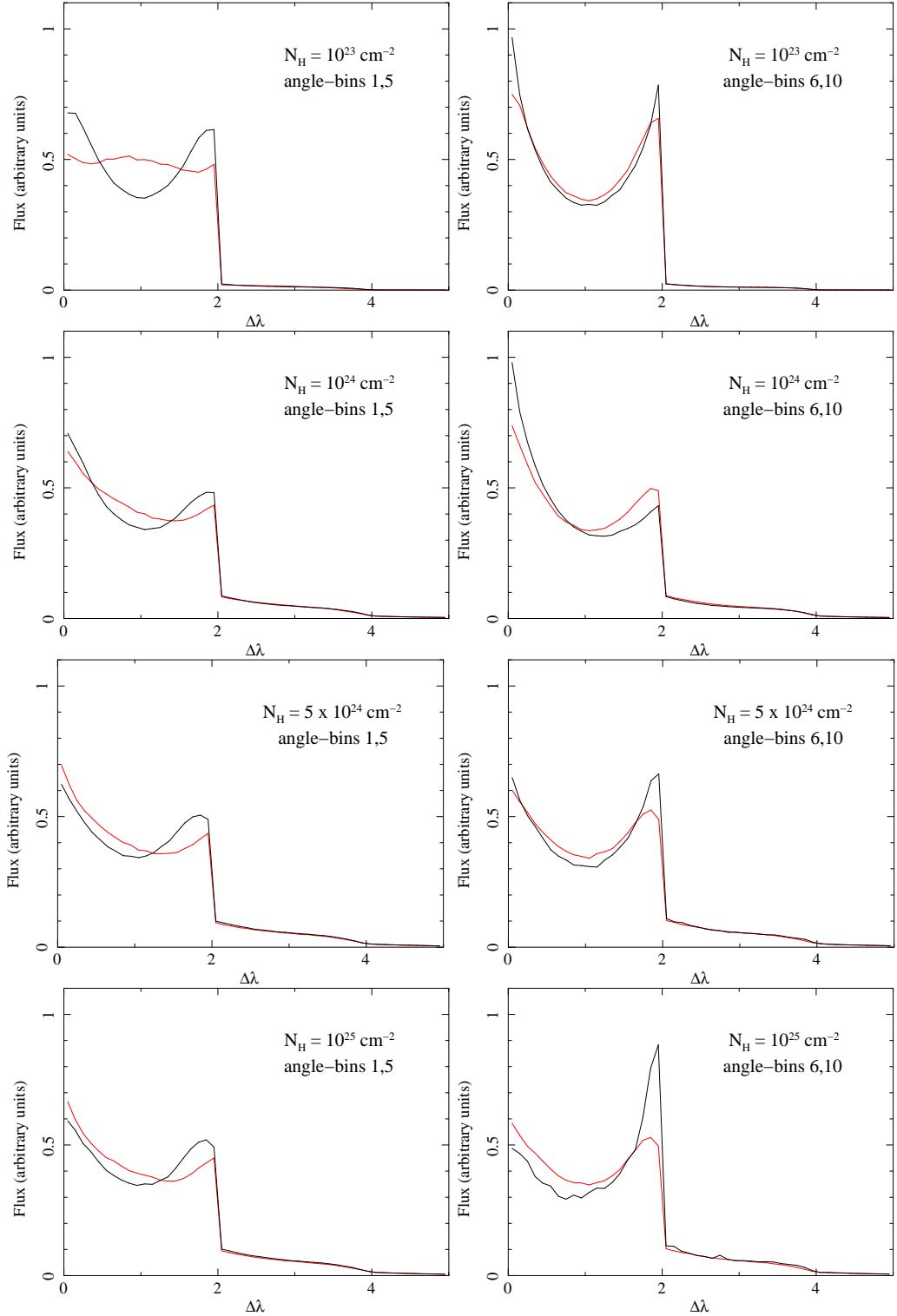

\vspace{-3mm}
\centerline{
\psfig{figure=f3a.ps,height=5.0cm,angle=270}
\psfig{figure=f3b.ps,height=5.0cm,angle=270}
}
\centerline{
\psfig{figure=f3c.ps,height=5.0cm,angle=270}
\psfig{figure=f3d.ps,height=5.0cm,angle=270}
}
\centerline{
\psfig{figure=f3e.ps,height=5.0cm,angle=270}
\psfig{figure=f3f.ps,height=5.0cm,angle=270}
}
\centerline{
\psfig{figure=f3g.ps,height=5.0cm,angle=270}
\psfig{figure=f3h.ps,height=5.0cm,angle=270}
}
\vspace{-2mm}
\caption[Shapes of the Compton shoulder for different column densities and inclination angles.]
{\footnotesize
The \fekalfa emission-line Compton shoulders for a
power-law
incident continuum with $\Gamma=1.9$, for four column densities
(from top to bottom,
$N_{\rm H} = 10^{23}$, $10^{24}$, $5 \times 10^{24}$, and
$10^{25} \rm \ cm^{-2}$) and different inclination angles of the torus.
The left-hand panels show the two extreme non-intercepting angle bins
\#1 (red) and \#5 (black), and the right-hand panels show the two extreme
intercepting angle bins, \#6 (red) and \#10 (black)-- see \tablecosrange
in MY09.
No velocity broadening has been applied.
{\it Note that in order to directly compare the Compton shoulder
shapes, the total flux for each shoulder has been renormalized to the same
value}. The line flux (in units of normalized flux per unit
wavelength shift) is plotted
against the dimensionless Compton
wavelength shift with respect to the zeroth-order rest-frame
energy of the emission line ($E_{0}$), $\Delta \lambda =
(511 \ {\rm keV}/E)-(511 \ {\rm keV}/E_{0})$.
}
\end{figure}

\begin{figure}
\vspace{-3mm}
\centerline{
\psfig{figure=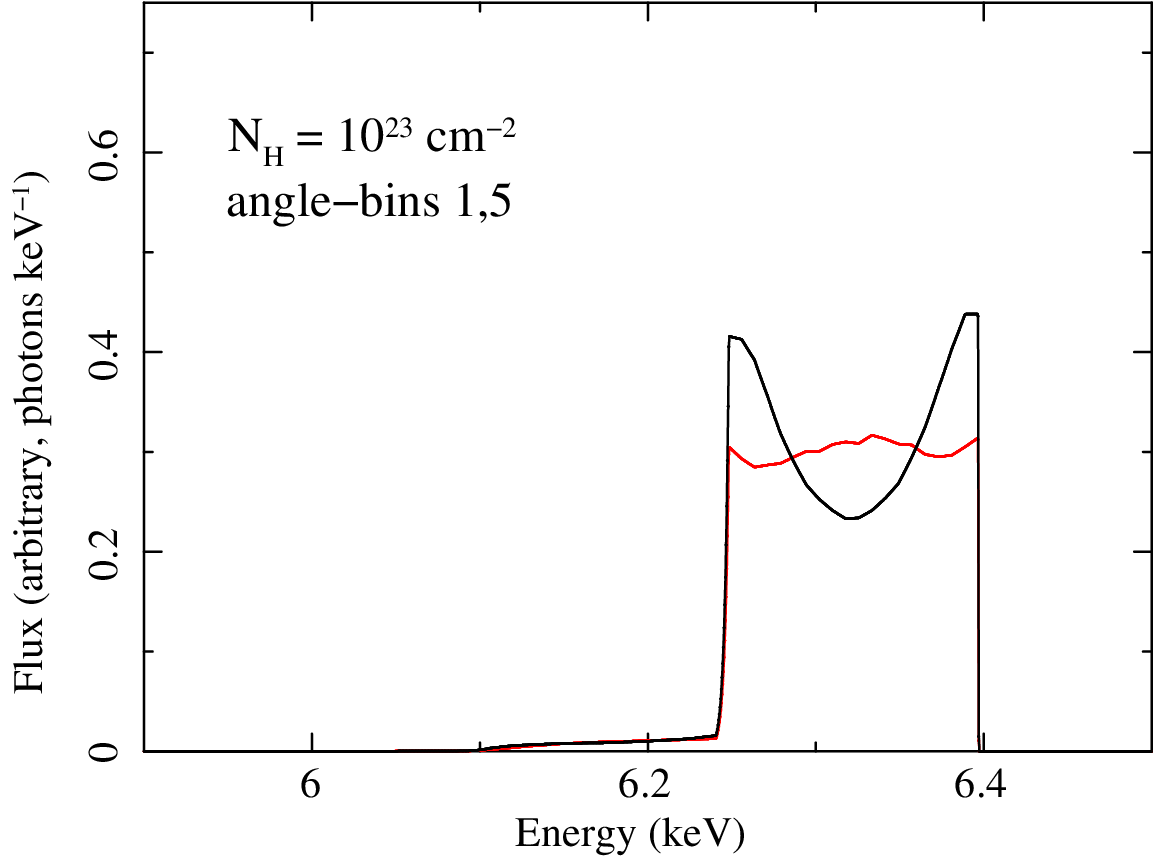,height=5.0cm,angle=270}
\psfig{figure=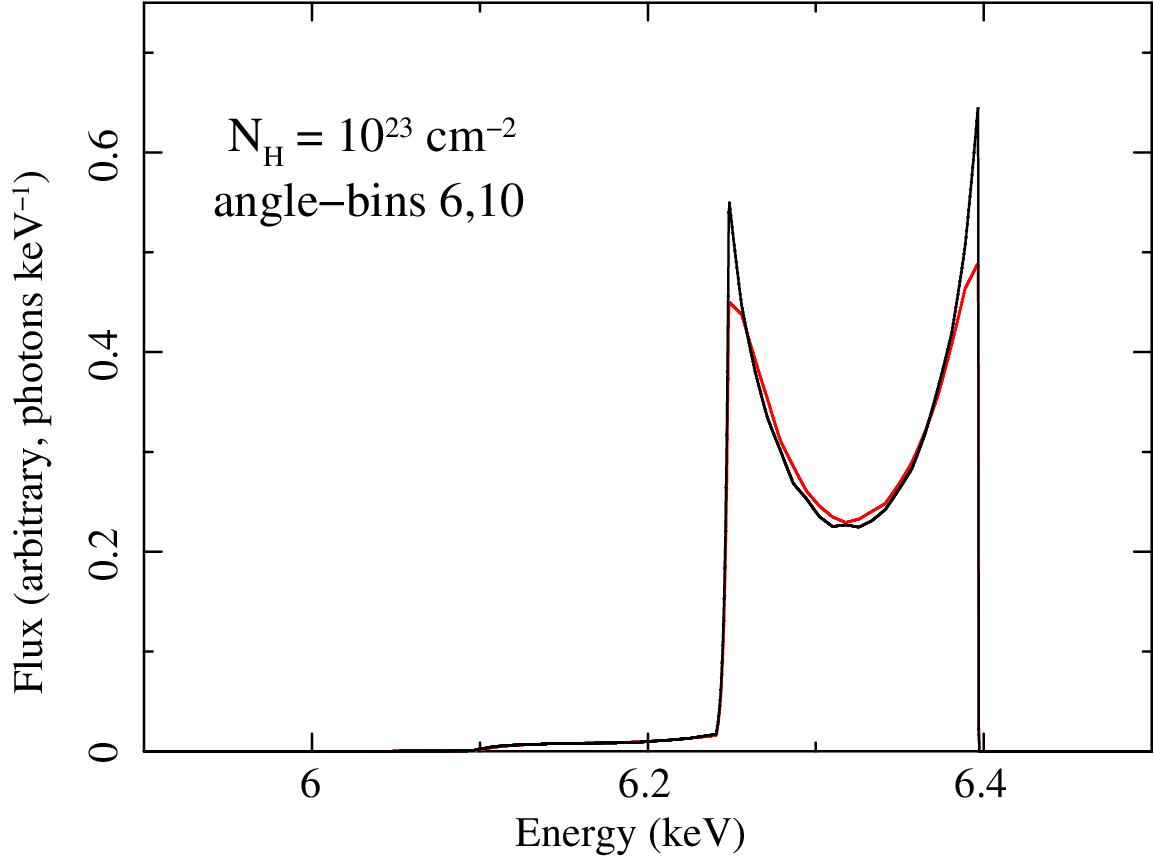,height=5.0cm,angle=270}
}
\centerline{
\psfig{figure=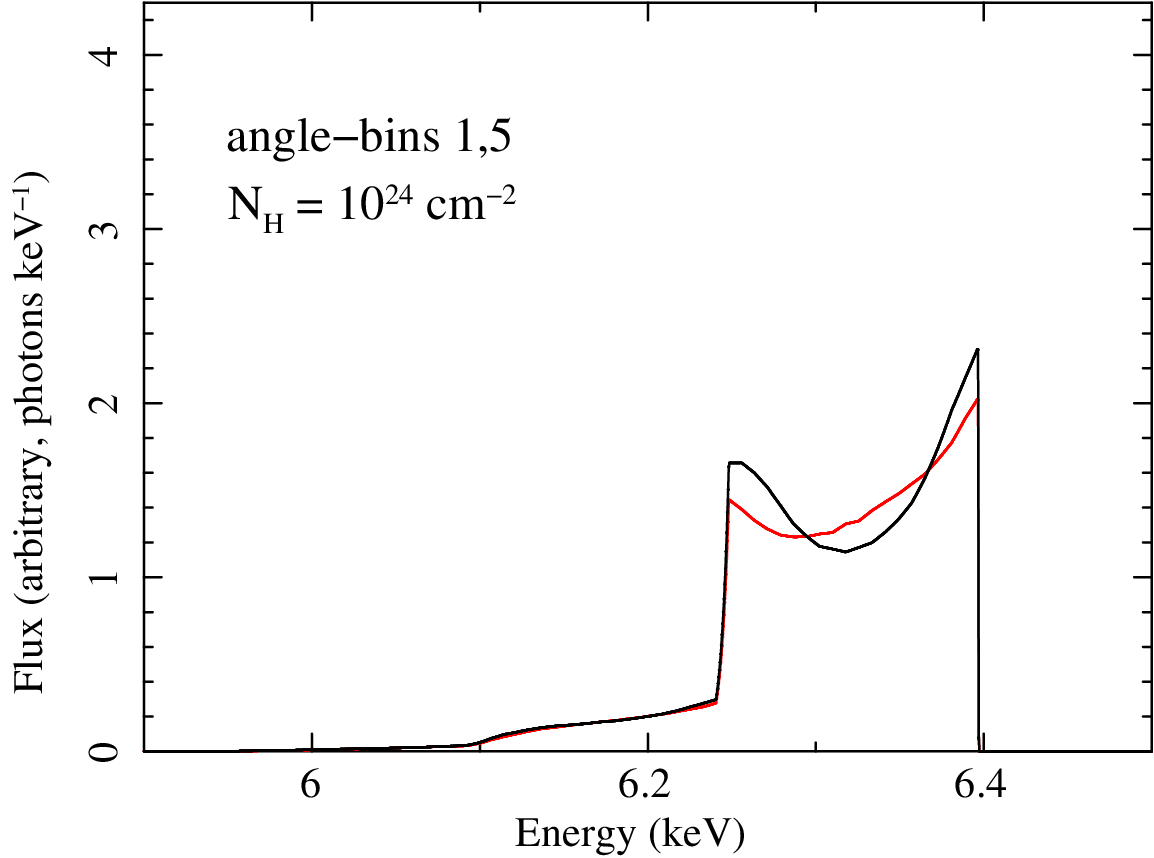,height=5.0cm,angle=270}
\psfig{figure=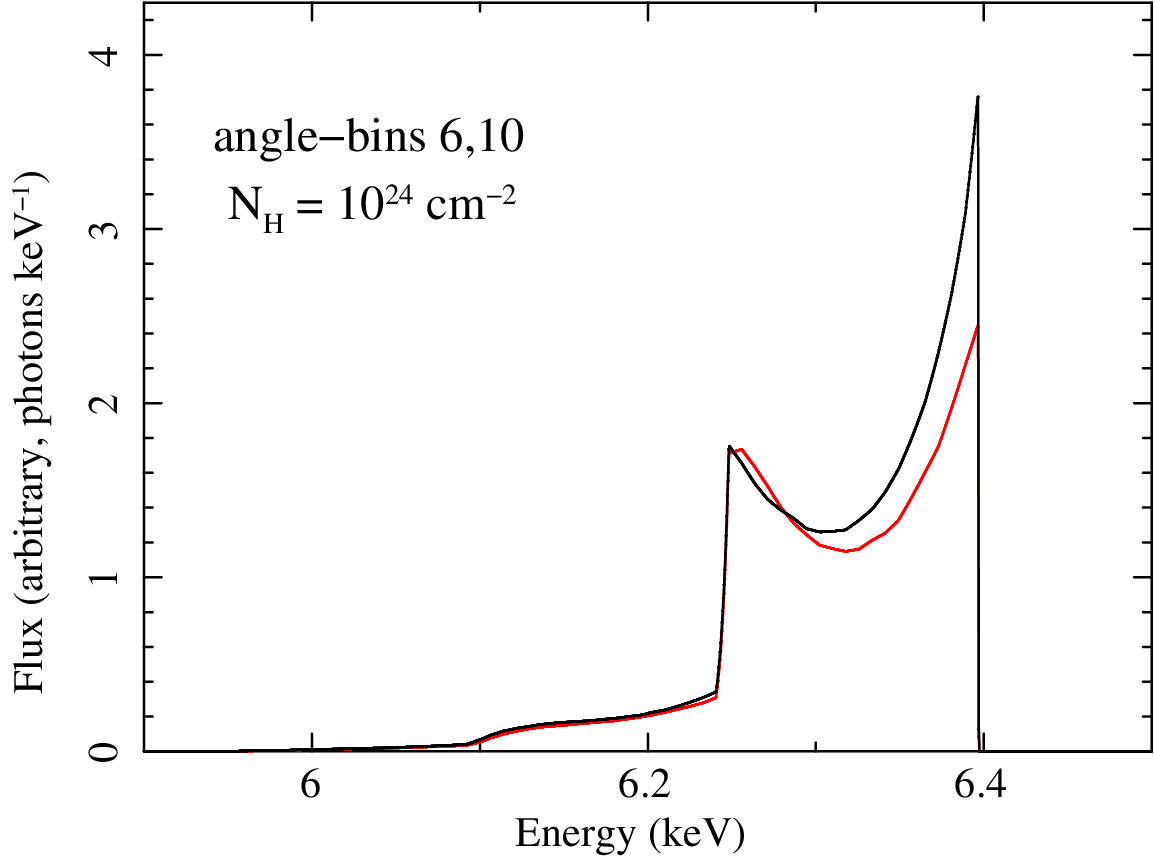,height=5.0cm,angle=270}
}
\centerline{
\psfig{figure=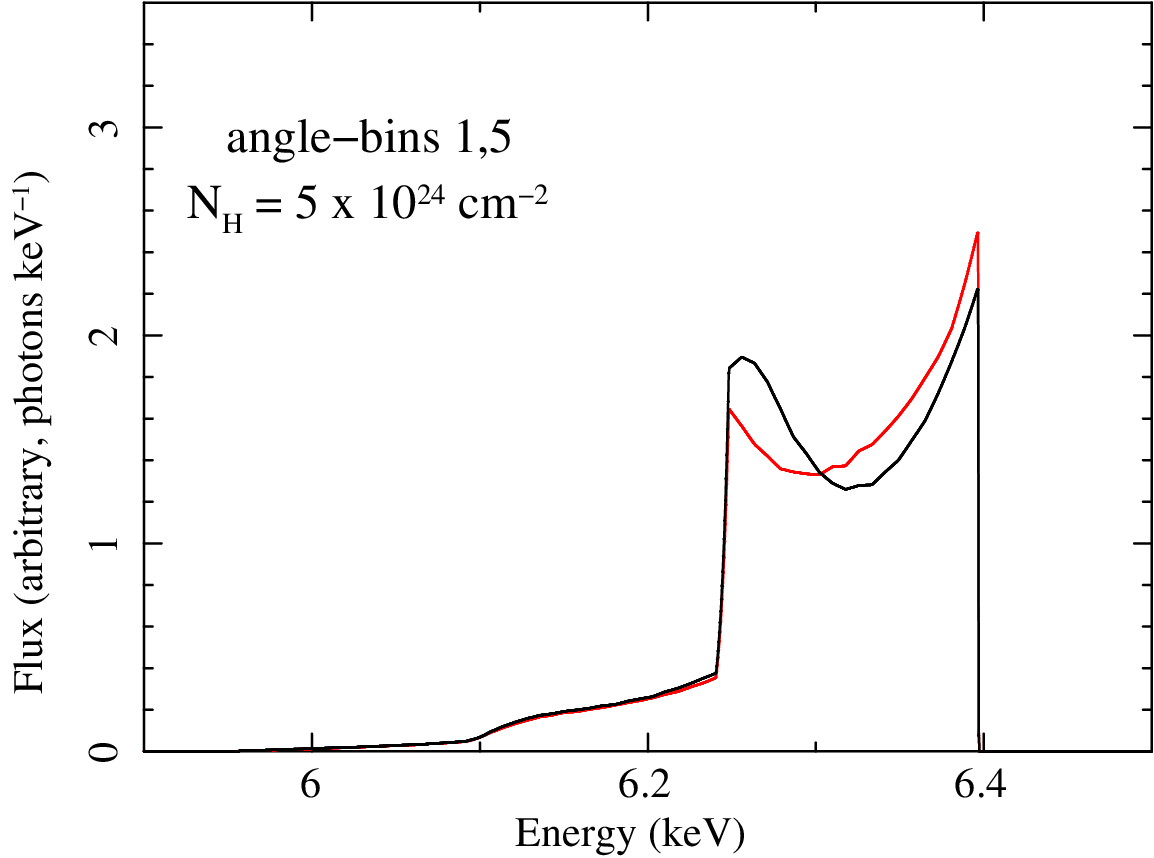,height=5.0cm,angle=270}
\psfig{figure=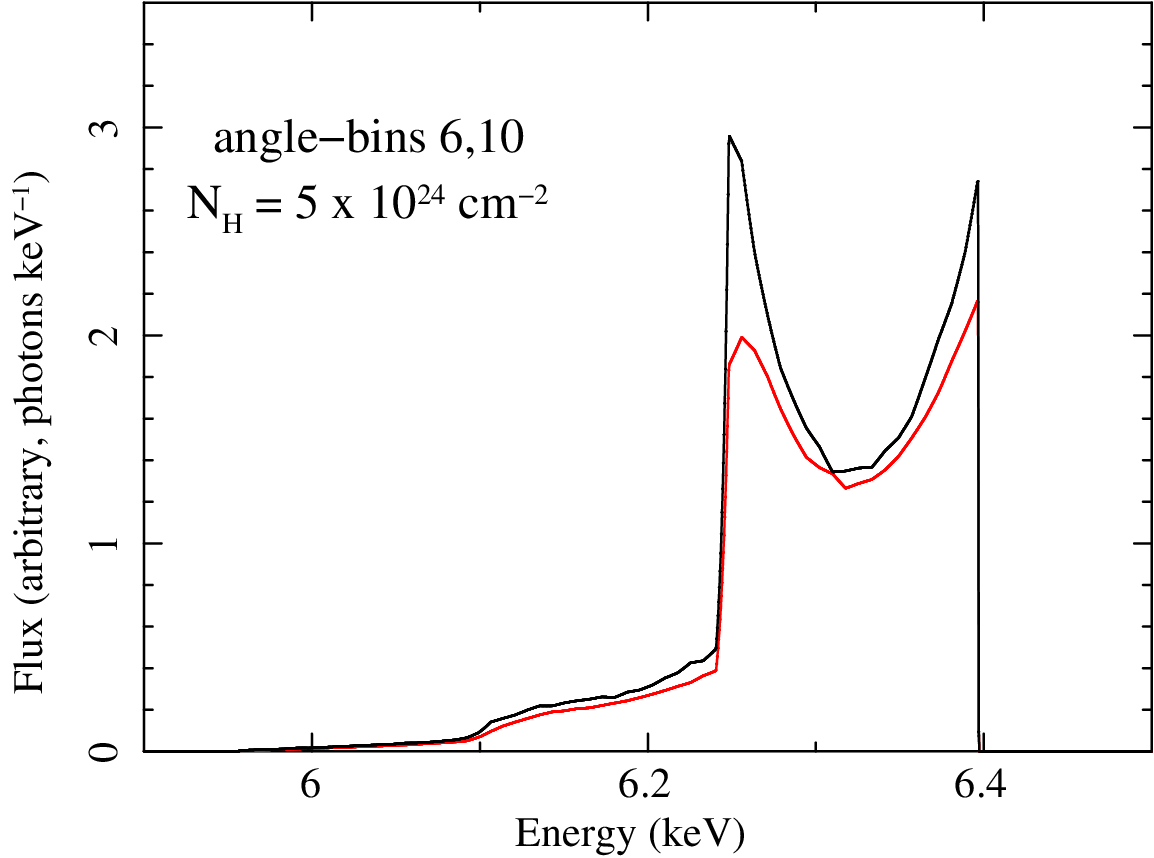,height=5.0cm,angle=270}
}
\centerline{
\psfig{figure=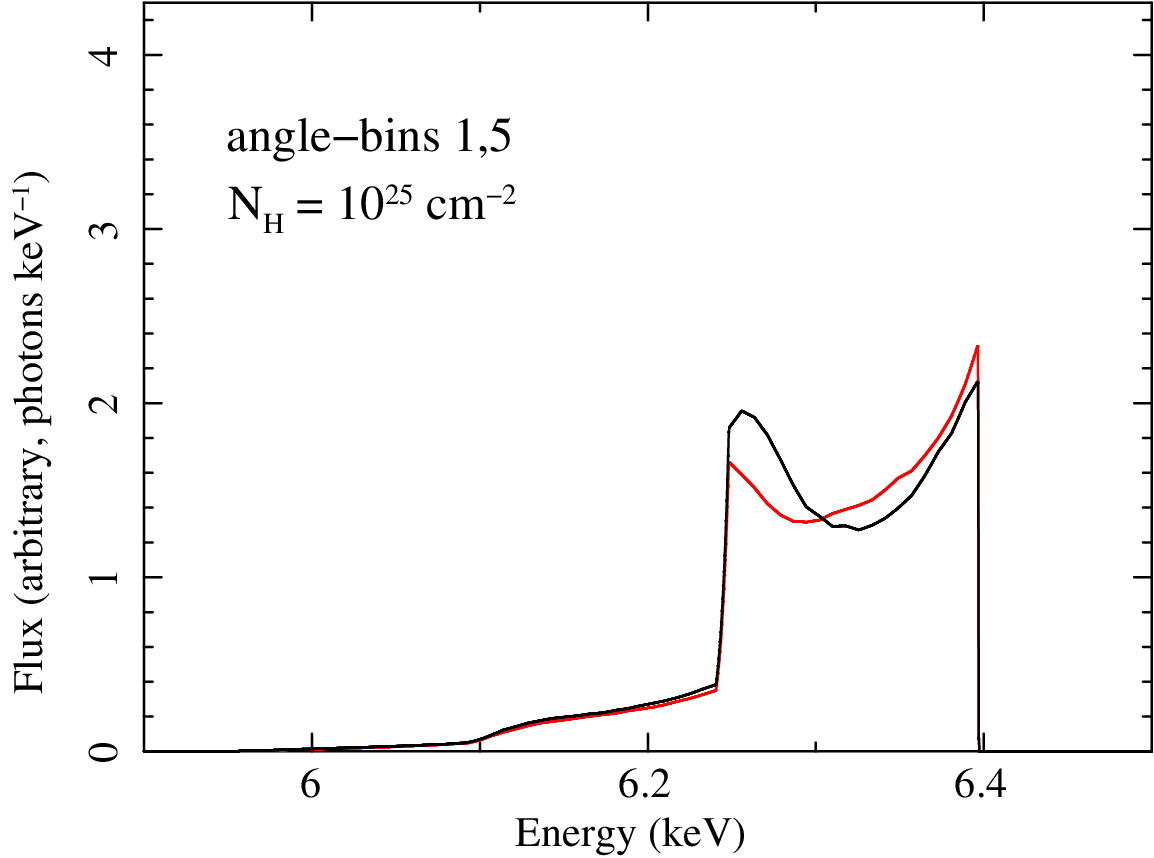,height=5.0cm,angle=270}
\psfig{figure=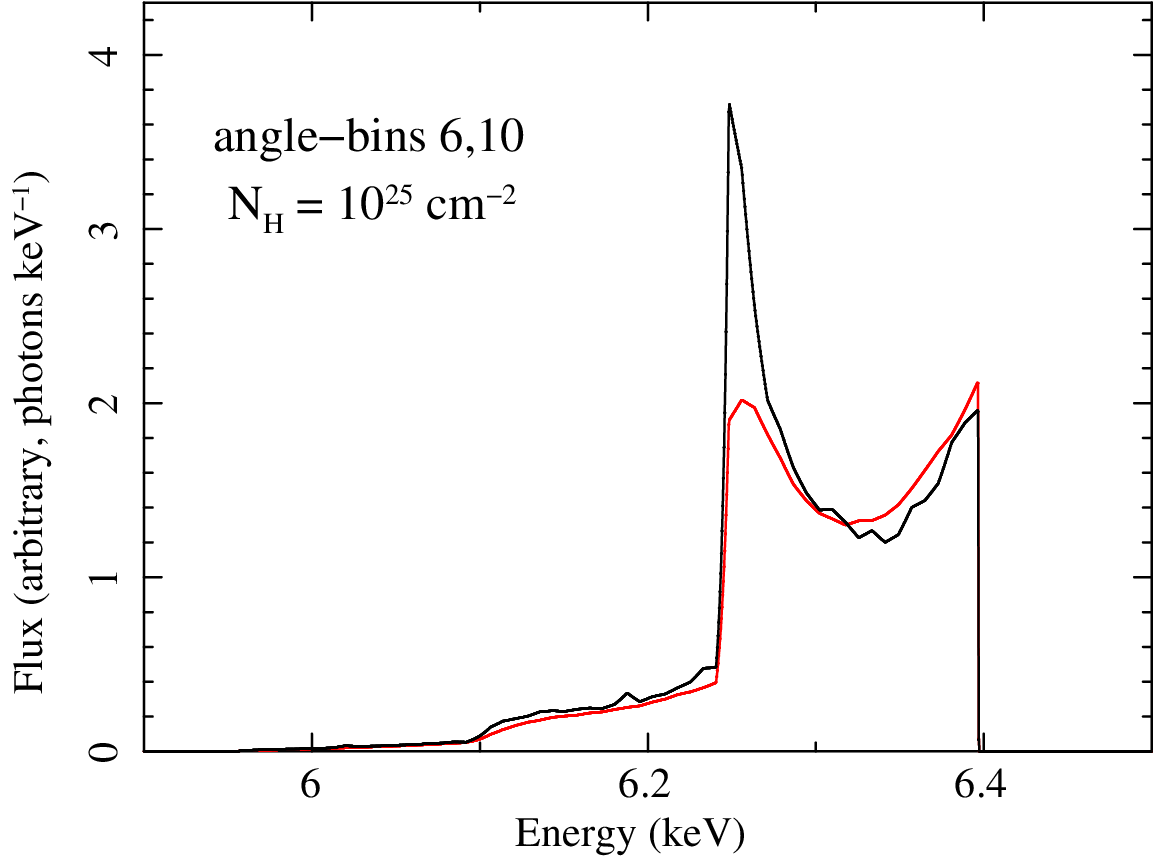,height=5.0cm,angle=270}
}
\vspace{-2mm}
\caption[Shapes of the Compton shoulder versus energy.]
{\footnotesize
The \fekalfa emission-line Compton shoulders for a
power-law
incident continuum with $\Gamma=1.9$, for four column densities
(from top to bottom,
$N_{\rm H} = 10^{23}$, $10^{24}$, $5 \times 10^{24}$, and
$10^{25} \rm \ cm^{-2}$) and different inclination angles of the torus.
The left-hand panels show the two extreme non-intercepting angle bins
\#1 (red) and \#5 (black), and the right-hand panels show the two extreme
intercepting angle bins, \#6 (red) and \#10 (black)-- see \tablecosrange
in MY09.
No velocity broadening has been applied.
The Compton shoulders correspond to identical parameters to
those in \figfekmcshouldersp, but here the profiles are plotted against
energy and the units of flux are $\rm photons \ \rm keV^{-1}$.
Although the overall absolute normalization is arbitrary, the Compton
shoulder profiles have relative normalizations that reflect
the original Monte Carlo results.
}
\end{figure}

Interpretation of \figfekmcshoulders is aided by the relation
between $\Delta\lambda$ and the scattering angle, $\alpha$:
$\Delta \lambda = 1-\cos{\alpha}$.
We can see from \figfekmcshoulders that at a column density of
$10^{23} \rm \ cm^{-2}$ the Compton shoulder is dominated by
the first scattering and the profile is determined by the
appropriate sampling of the single-scattering Thomson differential
cross-section (which is peaked for forward and backward scattering
and has a minimum for orthogonal scattering angles).
For the face-on Compton shoulder at such low column densities,
the forward and backward scattering peaks are suppressed, giving
an essentially flat Compton shoulder profile. This is because
most of the scattering occurs in the equatorial plane of the
torus when the scattering optical depth is small because
the equatorial plane presents the largest column density.
Thus, zeroth-order line photons with directions that roughly
align with the equatorial plane must scatter predominantly
at right angles in order to be observed in the face-on torus 
\thetaobs bin. On the other hand, those zeroth-order line photons that undergo 
forward and backward scattering are likely to be traveling 
in directions aligned with the equatorial plane again after 
scattering so are not as likely to appear in the face-on 
torus \thetaobs bin. For larger inclination angles, we see
in \figfekmcshoulders that the forward and backward scattering
peaks become more and more prominent, and for the edge-on
angle bin they are significantly enhanced and
the $90^{\circ}$ scattering direction is correspondingly suppressed.
We obtained similar Compton shoulder profiles for column densities
less than $10^{23} \rm \ cm^{-2}$.

As the column density of the torus increases, a scattered line photon
is more likely to escape the medium if it scatters near a surface.
Near the non-illuminated surfaces of the torus (i.e. the surfaces with
more material in between the X-ray source and the first scattering),
there are less zeroth-order \fekalfa line photons produced. 
The zeroth-order
line photons that do appear near the far-side
surfaces were more likely produced near an illuminated surface
and are therefore predominantly heading {\it towards} the far surface
rather than away from it.
However, even if a zeroth-order line photon is traveling towards
the surface (no matter how far from the X-ray source), 
if it is back-scattered, the scattered photon then 
travels back in to the medium, with a higher mean free path
to escape. Therefore, the scattered photons that are produced
by zeroth-order line photons traveling towards the surface {\it and}
undergo forward scattering will preferentially escape the medium.
We can see this in \figfekmcshoulders for $N_{H} = 10^{24} \rm \ cm^{-2}$,
which shows enhancement of the forward-scattering peak and
suppression of the back-scattering peak. The effect is more
and more
pronounced as the inclination angle becomes larger and
larger because back-scattered photons could be intercepted by
the other side of the torus, presenting a large mean free path
to escape compared to photons back-scattering near the surface
in directions perpendicular to the equatorial plane.
Also, the parts of the torus that scatter line photons into
non-intercepting lines of sight present a mean column density
to the X-ray continuum that is smaller than the equatorial
parts of the torus.
Thus, the edge-on Compton shoulder profiles for $N_{H} = 10^{24} \rm \ cm^{-2}$
are significantly asymmetric, weighted in favor of forward-scattering.

As the column density of the torus increases even further, the zeroth-order
\fekalfa line photons are created closer and closer to the
surfaces nearest the continuum illumination
since the continuum flux that produces the fluorescence
diminishes further into the medium.
Therefore scattered photons preferentially escape from the
illuminated surfaces.
Thus, the probability of scattered photons escaping from the medium
is not very different for forward-scattering and back-scattering
because now there is no selection effect for the original zeroth-order
line photon to be heading towards a surface before scattering.
At lower column densities, zeroth-order photons could end up
a longer way into the medium, far from an illuminated surface,
where it was less unlikely to find a zeroth-order photon traveling
{\it away} from the surface (in which case it would have to back-scatter
to escape). However, now for a higher column density, there
is a dearth of zeroth-order line photons near
surfaces that are not directly illuminated by the continuum
because large numbers are absorbed or scattered before they get
there, and the continuum is heavily extinguished, producing
less zeroth-order photons near far-side surfaces.
It can be seen from \figfekmcshoulders that for
$N_{H} = 5 \times 10^{24} \rm \ cm^{-2}$, the forward-scattering
peak is now indeed not particularly
strong compared to the back-scattering peak
and the Compton shoulder profiles 
have now become more symmetric again for the intercepting
lines of sight. Some of this
symmetry is caused by the higher-order scatterings
effectively raising the back-scattering peak. The Compton shoulder profiles
for the non-intercepting lines-of-sight are 
not very different to those for $N_{H} = 10^{24} \rm \ cm^{-2}$
(i.e they still have a slightly higher
weighting on the forward-scattering peak).
This is because for these scattered photons no material is being
added between the observer and the scattering sites
as the column density increases, and 
adding more optical depth deeper in the
medium does not make much difference because the deeper
scattering sites cannot make a further significant contribution to
the observed scattered flux. 

Now, moving on to the highest column density Compton shoulder
profiles in \figfekmcshoulders 
($10^{25} \rm \ cm^{-2}$) we see that those for the
non-intercepting lines-of-sight are essentially similar
to those for a column density of $5 \times 10^{24} \rm \ cm^{-2}$.
This is not surprising because the medium was already
Compton thick at that column density and the Compton shoulder
photons were already dominated by scattering near
the illuminated surfaces, with clear lines-of-sight to the observer.
However, the situation is different as we go to an edge-on
orientation of the torus. Here we see that it is the
{\it back-scattering} peak that now dominates over the forward-scattering
peak. The reason for this is that most of the torus
is so Compton-thick that all of the Compton shoulder
photons now come from two ``rings'' on 
opposite surfaces of the torus,
such that the rings
are regions having the smallest column density in directions parallel to the
equatorial plane. For zeroth-order line photons that are
created and scattered in this band of emission, there is 
no preference for forward and backward-scattering. However,
it is possible that zeroth-order line photons that 
escaped from the inner surface of the torus closer 
to the equatorial plane, can scatter in the ``ring'' and
be observed in the Compton shoulder profile. For the
smaller column density of $5 \times 10^{24} \rm \ cm^{-2}$
these photons could forward-scatter or backward-scatter in the
``ring'' and still reach the observer. However, at a column
density of $10^{25} \rm \ cm^{-2}$ the opacity is so high
for the forward-scattering route that it is suppressed. The
backward-scattering route on the other hand, does not have
to intercept the other side of the torus again before escape,
unless the line-of-sight is {\it exactly} edge-on. Recall that
our edge-on angle bin is actually a wedge that is a few degrees
wide (see \tablecosrange in MY09). Thus, for an exactly edge-on
orientation, the back-scattering peak would again be suppressed.
On the other hand, if the torus is actually patchy (for example,
consisting of a distribution of clouds), the back-scattering
peak would be observed even for exactly edge-on orientations.

The profile in \figfekmcshoulders (and in \figfekeshouldersp)
for the edge-on case for $N_{H} = 10^{25} \rm \ cm^{-2}$ is a {\it new},
type of Compton shoulder that has not been predicted before.
It does not appear in spherical or disk geometries (e.g. George \& Fabian 1991; Sunyaev \& Churazov 1996; Matt 2002), and 
to reveal it in a toroidal geometry requires
very high signal-to-noise Monte Carlo simulations, otherwise
it is not discernible from the noise. In energy space, this type of Compton shoulder
would show up in the X-ray spectrum as a peak at $\sim 6.24$~keV, 
separated from the zeroth-order \fekalfa line core at $\sim 6.4$~keV if the velocity
width is small enough.
If such a Compton shoulder were observed, it would be a powerful
indicator of a toroidal structure (possibly made of discrete clumps) with a column density of the
order of $10^{25} \rm \ cm^{-2}$, observed edge-on.
In contrast, a centrally-illuminated, uniform spherical geometry does not have 
any non-intercepting lines of sight so it can only produce Compton shoulder
profiles in the Compton-thick regime that are weighted towards forward
scattering. Moreover, an externally illuminated Compton-thick disk {\it only}
has non-intercepting lines of sight and can therefore only produce
essentially symmetric Compton shoulder profiles, with no preference
for forward or backward scattering.

\section{Dependence of the Compton Shoulder on the Illuminating Continuum}
\label{csgamdepend}

In \S\ref{fllcsratio} we reported a weak dependence of the 
ratio of the flux in the Compton shoulder to the zeroth-order \fekalfa line
flux on the photon index, $\Gamma$, of the power-law incident continuum.
Here we report the dependence of the {\it shape} of the Compton shoulder
profile on $\Gamma$.
\figfekcsdifgamm shows the Compton shoulder profiles for 
$N_{H} = 10^{25} \ \rm cm^{-2}$, directly comparing the cases
of $\Gamma=1.5$ (dotted curves) and $\Gamma=2.5$ (solid curves) for face-on ({\it left panel})
and edge-on ({\it right panel}) orientations. 
As in \figfekmcshoulders (and as described in \S\ref{fllcspardepend}),
the fluxes have been renormalized so that the
total flux is unity and the Compton
shoulder profiles are plotted against $\Delta \lambda$.
It can be seen that for the face-on orientation the
dependence of the shape of the Compton shoulder profile 
on the steepness of the incident X-ray continuum is extremely
weak and would not in practice be detectable. However, the
situation is different for the edge-on orientation.
The Compton shoulder profile is more symmetric for the
flatter incident spectrum ($\Gamma=1.5$), whilst the
steeper spectrum ($\Gamma=2.5$) still shows the back-scattering
peak. As discussed at the end of \S\ref{fllcspardepend}, for
column densities as high as $10^{25} \ \rm cm^{-2}$, two ``rings''
of emission on the surfaces of the torus that are parallel to the
equatorial plane contribute flux to the Compton shoulder profile
that has no forward or backward scattering preference. For
flatter incident continua, the \fekalfa line photons are formed
deeper in the medium because there are a relatively higher number
of high-energy continuum photons (than in steeper spectra), which can
penetrate deeper into the medium than lower energy photons.
The higher energy continuum photons also undergo
a larger number of mean scatterings than lower
energy before being absorbed by an Fe K edge so that the
resulting \fekalfa photons originate from a larger region.
For these reasons \fekalfa photons that result from the
absorption of high-energy continuum photons are more likely
to scatter locally before escape, but the \fekalfa photons
created by lower-energy continuum photons are more likely to
encounter their first scattering on the far side of the torus.  
Therefore the
Compton-shoulder has a relatively greater symmetric component
for $\Gamma=1.5$ than for $\Gamma=2.5$.

\begin{figure}
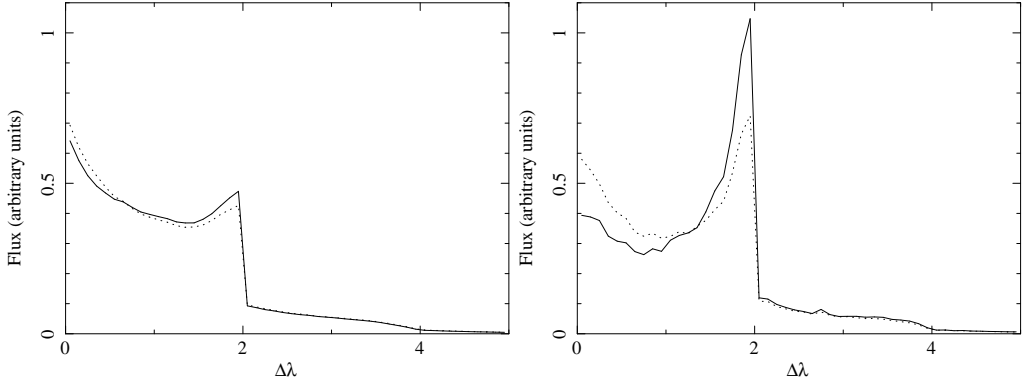

\centerline{
\psfig{figure=f5a.ps,height=5cm,angle=270}
\psfig{figure=f5b.ps,height=5cm,angle=270}
}
\caption[The dependence of the Compton shoulder shape on $\Gamma$.]
{\footnotesize
The \fekalfa emission-line Compton shoulders for
$N_{\rm H} =10^{25} \rm \ cm^{-2}$ and
a power-law
incident continuum with
$\Gamma=1.5$ (dotted), and $\Gamma=2.5$ (solid).
The {\it left-hand} panel corresponds to the face-on angle bin and the
{\it right-hand}
panel corresponds to the edge-on angle bin.
No velocity broadening has been applied.
{\it Note that in order to directly compare the Compton shoulder
shapes, the total flux for each shoulder has been renormalized to the same
value}. The line flux (in units of normalized flux per unit
wavelength shift) is plotted
against the Compton
wavelength shift with respect to the zeroth-order rest-frame
energy of the emission line ($E_{0}$), $\Delta \lambda =
(511 \ {\rm keV}/E)-(511 \ {\rm keV}/E_{0})$.
}
\end{figure}

\section{Velocity Broadening}
\label{fllvelbroad}

\begin{figure}
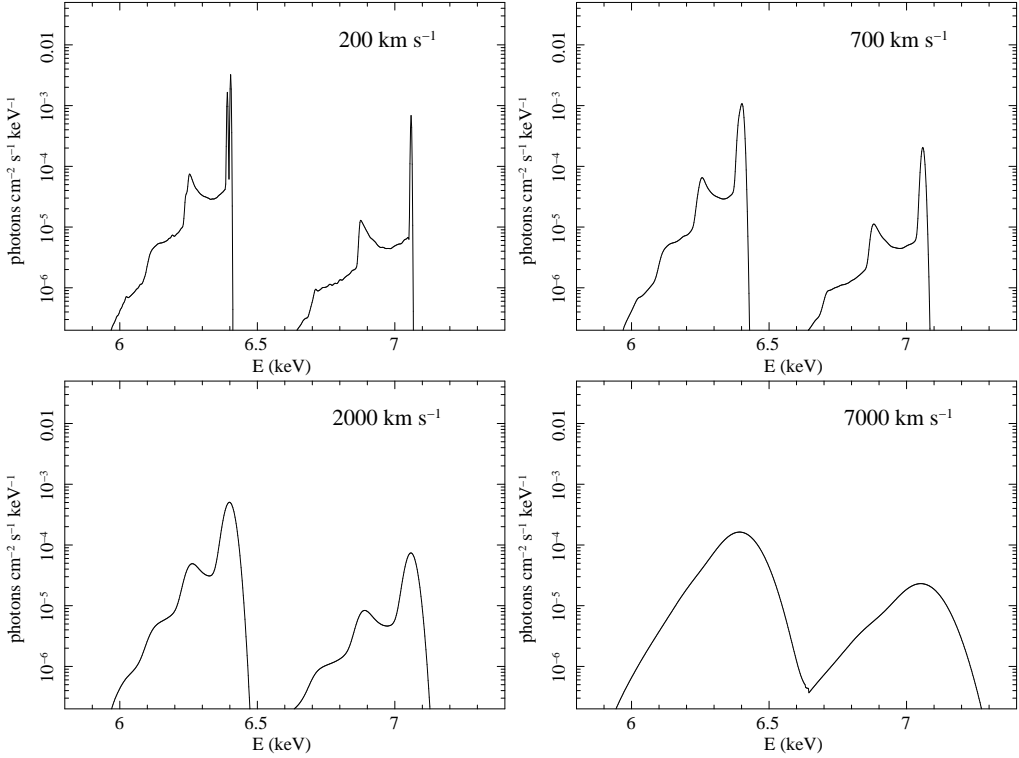

\centerline{
\psfig{figure=f6a.ps,height=5cm,angle=270}
\psfig{figure=f6b.ps,height=5cm,angle=270}
}
\centerline{
\psfig{figure=f6c.ps,height=5cm,angle=270}
\psfig{figure=f6d.ps,height=5cm,angle=270}
}
\caption[Effect of velocity broadening on the emission-line spectrum
(\nh $=10^{25} \ \rm cm^{-2}$, edge-on).]
{\footnotesize Effect of velocity broadening
on the fluorescent emission-line spectra consisting
of the Fe~K$\alpha_{1}$, Fe~K$\alpha_{2}$, and \fekbeta lines.
A Gaussian convolution function was applied to the
\mchyphen results for a toroidal X-ray reprocessor
with $N_{\rm H} = 10^{25} \ \rm cm^{-2}$ (viewed {\it edge-on}),
illuminated by a power-law continuum with a photon index
of $\Gamma=1.9$.
Four emission-line spectra are shown with a FWHM
of 200, 700, 2000, and 7000~$\rm km \ s^{-1}$.
{\it Note the logarithmic flux axis: the narrower core of the line
for the smaller velocities makes the flux per keV in the core much
higher than that in the Compton shoulder}.
}
\end{figure}

So far we have discussed the Compton shoulder profiles
calculated directly from the Monte Carlo simulations, which do not
include any kinematics. In practice, Doppler shifts due to bulk motion
will broaden the Compton shoulders and such velocity broadening should
be included when fitting real data with a model of the \fekalfa line. 
We also wish to study the possible confusion in velocity width measurements resulting from
the doublet-splitting
of the \fekalfa line. This is going to be especially important
for observations of AGN made with calorimeters (as planned for \astrohp) which
will have a spectral resolution for the \fekalfa line of
several eV (for ``local'', low-redshift AGN).
For neutral Fe, the K$\alpha$ emission consists of two lines,
K$\alpha_{1}$ at 6.404 keV and K$\alpha_{2}$ at 6.391 keV, with a
branching ratio of 2:1 (e.g. see
Bambynek \etal 1972). In the \mchyphen
simulations these were
treated as a single line with an energy of 6.4008~keV
(corresponding to the weighted mean
energies of the Fe~K$\alpha_{1}$ and Fe~K$\alpha_{2}$ lines).
For the \mytorus spectral-fitting model, the two components of
the \fekalfa line are {\it reconstructed} using the above
branching ratio and line energies.
The energy difference between the two components
is small enough to neglect the differences in opacities and
the error incurred in this procedure is too small to impact
fitting even \astroh data.
For each of the Fe line components, K$\alpha_{1}$ and Fe~K$\alpha_{2}$,
we can construct a Compton shoulder for a given set of torus parameters,
and sum the two profiles with the correct branching ratio.
We then apply simple velocity broadening by convolving the
summed profile with a Gaussian. The actual velocity-broadening
function is {\it unlikely} to be a Gaussian, but
it is also unlikely that even \astroh will
be able to distinguish 
between different geometries of the distant-matter \fekalfa
line in AGN just from its velocity profile (e.g. see Yaqoob \etal 1993).

The effect of velocity broadening on the \fekalfa fluorescent line
spectrum is illustrated in \figfekvbroadfour
for $N_{\rm H} = 10^{25} \ \rm cm^{-2}$ and the edge-on orientation
(angle bin \#10).
This time {\it we include the zeroth-order} components of the emission
lines, as well as the Compton shoulders, and we plot the spectra
as a function of energy (and not $\Delta \lambda$).
We also show the \fekbeta emission line which is also produced
by our Monte Carlo simulations (see MY09 for details). 
A simple Gaussian function has been employed for convolving the
emission-line profiles, using a FWHM velocity that is constant
with respect to energy. A total of four components are 
thus convolved (the zeroth-order components of Fe~K$\alpha_{1}$ and
 Fe~K$\alpha_{2}$, and their respective Compton shoulders).
Spectra are shown for four values
of the FWHM, namely $200 \rm \ km \ s^{-1}$,
$700 \rm \ km \ s^{-1}$, $2000 \rm \ km \ s^{-1}$, and
$7000 \rm \ km \ s^{-1}$.
The value of $200 \rm \ km \ s^{-1}$ is approximately
the velocity resolution that will be achieved in
the Fe~K band by the calorimeters
aboard \astrohp.
The value of $700 \rm \ km \ s^{-1}$ is in the regime
expected for the classical, parsec-scale torus.
The value of $2000 \rm \ km \ s^{-1}$ is approximately the
velocity resolution of the \chandra high energy grating (HEG)
spectrometer in the Fe~K energy band. It is also typical
of optical emission-line widths in the outer BLR.
The value of $7000 \rm \ km \ s^{-1}$ is approximately
the velocity resolution of CCD detectors in the Fe~K band.
All emission-line spectra in \figfekvbroadfour
were calculated for an
incident power-law spectrum with $\Gamma=1.9$.

We see from \figfekvbroadfour that the blending
that results from velocity broadening can significantly
dilute the characteristic
features of the Compton shoulder and the combined \fekalfa line profile.
Notice that the zeroth-order cores
for the Fe~K$\alpha_{1}$ and Fe~K$\alpha_{2}$ components
cannot be distinguished as separate components
even for a FWHM as low as $700
\ \rm km \ s^{-1}$
(this has already
been pointed out in Yaqoob \etal 2001).
At a FWHM of $2000 \rm \ km \ s^{-1}$ the Compton shoulder
and line core are still resolved and 
the back-scattering peak is still resolved, although already
heavily blended over the width of the shoulder. 
At a FWHM of
$7000 \rm \ km \ s^{-1}$ {\it there is no shoulder to the
line profile} and the only trace of it is a slight asymmetry
in the line profile.
An obvious implication of this is that
{\it the Compton shoulder cannot be unambiguously
resolved with CCD detectors}.

\section{Summary}
\label{summary}

We have studied the properties of the
Compton shoulder of the \fekalfa emission line in the toroidal X-ray 
reprocessor model of Murphy \& Yaqoob (2009)
for equatorial column densities in the
range $N_{H}=10^{22} \ \rm cm^{-2}$ to $10^{25} \ \rm cm^{-2}$
for the case of a half-opening angle of $60^{\circ}$ and
the cosmic abundances of Anders \& Grevesse (1989).
The shape of the Compton shoulder and its flux relative to the 
core of the \fekalfa emission line provide important physical constraints
on the line-emitting matter, including column density and orientation of
the reprocessing structure. We have made our results available
in a form (the \mytorus model)
that is suitable for direct spectral-fitting to real,
high spectral-resolution data. 

We find that the ratio of the flux in the Compton shoulder
to that in the unscattered core component of the \fekalfa line
increases approximately in proportion to $N_{H}$ until
$N_{H} \sim 5 \times 10^{23} \ \rm cm^{-2}$. The ratio reaches a
maximum value that depends on the orientation of the torus
($\sim 0.29$ for face-on and $\sim 0.37$ edge-on), at 
$N_{H} \sim 2-3 \times 10^{24} \ \rm cm^{-2}$. 
The toroidal geometry gives a wider
variety of Compton shoulder
profiles than either (centrally-illuminated) spherical 
or disk geometries. 
When the torus is observed with non-intercepting lines-of-sight,
the shape of the Compton shoulder of the \fekalfa line
as a function of Compton wavelength shift is approximately
symmetrical over the wavelength interval corresponding to the
first scattering. For intercepting lines-of-sight this
symmetry is present in the optically-thin limit but the
Compton shoulder profile becomes weighted more and more towards
the zero wavelength shift side of the profile 
as the column density increases, 
even to only $N_{H} \sim 10^{23} \ \rm cm^{-2}$. The asymmetry
continues to increase up to $N_{H} \sim 2-3 \times 10^{24} \ \rm cm^{-2}$.
However, by $N_{H} \sim 5 \times 10^{24} \ \rm cm^{-2}$ the Compton
shoulder (in the first scattering interval) becomes symmetrical
again.

Our Monte Carlo simulations have
been performed with high statistical accuracy and
reveal that the case of an edge-on, Compton-thick 
torus produces a new type of Compton shoulder that is unique to
the toroidal geometry. It is dominated by a
narrow back-scattering peak at $\sim 6.24$~keV.
In addition, our results are sensitive
enough to reveal a weak dependence of the shape of the Compton shoulder
and its magnitude (relative to the \fekalfa line core), on the 
spectral shape of
the incident X-ray continuum.
We have also presented results on the effect of 
velocity broadening on the \fekalfa line
profile and find that if either the velocity width or instrument resolution
is greater than a FWHM of $\sim 2000 \ \rm km \ s^{-1}$, the Compton shoulder
begins to become blended with the line core and the characteristic features
of the Compton shoulder become harder to resolve. In particular, at 
a FWHM of $7000 \ \rm km \ s^{-1}$ the Compton shoulder is not resolved
at all, its only signature being a weak asymmetry in the blended line profile.
This means that CCD X-ray detectors cannot 
unambiguously resolve the Compton shoulder.

The properties of the Compton shoulder as a function of the
torus opening angle (or covering factor) and the abundance of Fe cannot
be trivially deduced and require much more extensive Monte Carlo 
simulations. This will be the subject of future work.

Acknowledgments \\
Partial support (TY) for this work was provided by NASA through \chandra Award
TM0-11009X, issued by the Chandra X-ray Observatory Center,
which is operated by the Smithsonian Astrophysical Observatory for and
on behalf of the NASA under contract NAS8-39073.
Partial support from NASA grants NNX09AD01G and NNX10AE83G is also
(TY) acknowledged.

\end{document}